\documentclass{aa}

\usepackage{graphics}
\usepackage{psfig}

\newcommand{\Hd}{\ensuremath{\mathrm H_2}}
\newcommand{\nHd}{$n_{\Hd}$}
\newcommand{\NHd}{$N_{\Hd}$}
\newcommand{\CuoO}{\ensuremath{\mathrm C^{18} \mathrm O}}  
\newcommand{\NCuoO}{$N_{\CuoO}$}
\newcommand{\utCO}{\ensuremath{^{13}\mathrm{CO}}}
\newcommand{\NutCO}{$N_{\utCO}$}
\def\NHt{NH$_3$}
\newcommand{\TK}{$T_{\mathrm K}$}
\newcommand{\juce}{\mbox{$J=1\rightarrow0$}}
\newcommand{\jdu}{\mbox{$J=2 \rightarrow 1$}}
\newcommand{\jdce}{\mbox{$J=2 \rightarrow 0$}}
\newcommand{\jtu}{\mbox{$J=3 \rightarrow 1$}}
\def\kms{~km~s$^{-1}$}
\def\cmmt{~cm$^{-3}$}
\def\cmmd{~cm$^{-2}$}
\def\smu{s$^{-1}$}
\newcommand{\ns}{M$-0.96+0.13$}
\newcommand{\cc}{M$-0.55-0.05$}
\newcommand{\cce}{M$-0.50-0.03$}
\newcommand{\qd}{M$-0.42+0.01$}
\def\td{M$-0.32-0.19$}
\def\uc{M$-0.15-0.07$}
\def\us{M$+0.16-0.10$}
\def\du{M$+0.21-0.12$}
\def\dq{M$+0.24+0.02$}
\def\tc{M$+0.35-0.06$}
\def\qo{M$+0.48+0.03$}
\def\co{M$+0.58-0.13$}
\def\sis{M$+0.76-0.05$}
\def\ot{M$+0.83-0.10$}
\def\nq{M$+0.94-0.36$}

\def\dnn{M$+2.99-0.06$}

\newcommand{\gsim}{\raisebox{-.4ex}{$\stackrel{>}{\scriptstyle \sim}$}}
\newcommand{\lsim}{\raisebox{-.4ex}{$\stackrel{<}{\scriptstyle \sim}$}} 
\def\le{$\leq$}

\begin{document}

   \thesaurus{08(10.03.1; 13.09.3; 13.09.4; 09.13.2; 09.03.1; 09.04.1)}

   \title{Warm H$_2$ in the  Galactic center region            
          \thanks{ Based on observations with ISO, an ESA project 
	        with instruments funded by ESA Member States (especially 
	        the PI countries: France, Germany, the Netherlands and 
            the United Kingdom) and with the participation of ISAS and NASA.}
         }


   \author{N.J.~Rodr\'{\i}guez-Fern\'andez\inst{1}, 
           J.~Mart\'{\i}n-Pintado\inst{1}, A.~Fuente\inst{1}, 
	   P.~de~Vicente\inst{1}, T.L.~Wilson\inst{2,3}, 
	   and S.~H\"uttemeister\inst{4}}

   \offprints{N.J.~Rodr\'{\i}guez-Fern\'andez}

   \institute{Observatorio Astron\'omico Nacional, IGN,
        Apartado 1143, E-28800 Alcal\'a de Henares, Spain; 
        nemesio$@$oan.es
   \and Max-Planck-Institut f\"ur Radioastronomie, Postfach 2024, 
	D-53010 Bonn, Germany
   \and Sub-mm Telescope Observatory, Steward Observatory, The University
          of Arizona, Tucson, AZ 85728, USA 
   \and Radioastronomisches Institute der Universit\"at Bonn, Auf dem
	H\"ugel 71, D-53121 Bonn, Germany  }

   \date{Received; accepted }

   \authorrunning{Rodr\'{\i}guez-Fern\'andez et al. }
   \titlerunning{Warm \Hd\ in the Galactic center}
   \maketitle

   \begin{abstract}
We present ISO observations of several \Hd\ pure-rotational lines
(from  S(0) to S(5)) towards a sample of 16 molecular
clouds distributed along the central $\sim 500$ pc of the Galaxy.
We also present \CuoO\ and \utCO\  \juce\ and \jdu\ observations
of these sources made  with the IRAM-30m telescope.
With the CO data we derive \Hd\ densities of 10$^{3.5-4.0}$~\cmmt\ and  \Hd\
column densities of a few 10$^{22}$~\cmmd.
We have  corrected the \Hd~ data 
for $\sim 30$ magnitudes of visual extinction 
using a self-consistent method. 
In every source, we find that the \Hd\ emission   exhibits a large
temperature gradient.
The S(0) and S(1) lines trace temperatures ($T$)  of $\sim 150$~K while the
S(4) and S(5) lines indicate temperatures of $\sim 600$~K.
The warm \Hd\ column density is  typically 
$\sim 1-2\times 10^{22}$~\cmmd, and is predominantly gas with $T$=150 K. 
This is the first direct  estimate of the total column 
density of the warm molecular gas in the Galactic center region.
These warm \Hd\ column densities represent a fraction of 
$\sim 30 \%$ of the gas traced by  the CO isotopes emission.
The cooling by \Hd~ in the warm component is comparable to that by CO.
Comparing our \Hd\ and CO data with available ammonia (\NHt) observations
from literature one obtains relatively high \NHt\ abundances of
a few $10^{-7}$ in both the warm and the cold gas.
A single shock or 
Photo-Dissociation Region (PDR)  cannot explain  all the observed \Hd\ lines.
Alternatives for the heating mechanisms are discussed.

      \keywords{ISM: clouds -- ISM: molecules -- ISM: dust, extinction --
                Galaxy: center -- Infrared: ISM: continuum -- 
                Infrared: ISM: lines and bands}
   \end{abstract}

%

\section{Introduction}
The interstellar matter in the inner few hundreds parsecs
of the Galaxy  (hereafter GC)  is mainly molecular.
In this region there are  molecular clouds  and
huge cloud complexes like Sgr~B2 which can be as
large as 70 pc, with masses of  10$^6$ solar masses.
The physical conditions in the GC clouds differ
appreciably to those of the clouds of the disk of the Galaxy.
The GC clouds have average densities of $\sim 10^4$ \cmmt\ instead
of 10$^2$ \cmmt\ typical of the clouds of the disk.
In addition, with widespread high temperatures (up to 200 K),
GC clouds are hotter than disk clouds. 

The temperatures of the warm  gas are known mainly by observations
of ammonia (\NHt ) metastable lines. 
G\"usten et al. (\cite{Gusten81}, \cite{Gusten85}) derived rotational 
temperatures ($T_\mathrm{rot}$) of 60-120 K in several GC clouds,
most of them in the Sgr A complex.
Morris et al. (\cite{Morris83}) showed that $T_\mathrm{rot} \sim 30-60$~K
are common in the region $|l|<2^\circ$.
The most complete study of the temperature structure of the molecular
gas in the GC, was carried out by
H\"uttemeister et al. (\cite{Huttemeister93}).
They presented a multilevel
study of \NHt\ metastable lines  of 36
molecular clouds distributed all along the ``Central Molecular Zone" (CMZ,
in notation of Morris \& Serabyn \cite{MS96}) and the ``Clump 2" complex,
which, although not belonging to the actual CMZ, exhibits similar properties.
They detected warm gas at all galactic longitudes and showed that
the \NHt\ emission can be characterized by two temperature components since
the $T_\mathrm{rot}$ derived from the (1,1) and (2,2) levels is $\sim 20-30$~K
and that derived from the (4,4) and (5,5) levels is  $\sim$ 70--200~K.
Unfortunately, the {\it a priori} unknown abundance of the \NHt\ molecule
has made it difficult to estimate the total column density of warm gas
in the GC clouds. 

The heating of the molecular gas over large regions ($\sim 10$~pc)
where the dust temperature is lower than 30 K
(Odenwald \& Fazio \cite{OF84}, Cox \& Laureijs \cite{CL89},
Mart\'{\i}n-Pintado et al. \cite{MP99}, Rodr\'{\i}guez-Fern\'andez
et al. \cite{RF00}) is a puzzle.
Indirect arguments such as  the large widths of molecular lines
or large abundances
in gas phase of molecules such as SiO (Mart\'{\i}n-Pintado et al. \cite{MP97},
H\"uttemeister et al. \cite{Huttemeister98})
or \NHt\ points towards a mechanical heating.
Wilson et al. (\cite{Wilson82}) proposed the dissipation of
turbulence induced by differential Galactic rotation as a  
possible heating source.

For the first time, we have  measured the total  column densities of warm gas
in the GC clouds by observing the lowest \Hd\ pure-rotational
transitions with the {\em Infrared Space Observatory} (ISO; Kessler et al.
\cite{Kessler96}).
The \Hd\ pure-rotational lines trace gas with temperatures of a few
hundred K (see Shull $\&$ Beckwith \cite{SB82} for a review on the
properties and the notation of the \Hd\ molecule).
ISO has  detected  \Hd\ pure-rotational lines in a variety of sources
such as: Young Stellar Objects (Van den Ancker \cite{vdA99}); galactic nuclei
(see e.g. Kunze et al. \cite{Kunze99});
Photo Dissociation Regions (PDRs) like NGC 7023
(Fuente et al. \cite{Fuente99}, \cite{Fuente00}) or S140
(Timmermann et al. \cite{Timmermann96});
shock-excited sources such as Orion Peak 1
(Rosenthal et al. \cite{Rosenthal00});
and proposed x-ray excited regions (XDRs) like RCW 103
(see Wright \cite{Wright00}). 

Our sample consists    of 18 molecular clouds from the samples
of H\"uttemeister et al. (1993) and Mart\'{\i}n-Pintado et al. (1997).
Two of these show a non-equilibrium \Hd\ ortho-to-para ratio and
have been studied in detail  by Rodr\'{\i}guez-Fern\'andez et al. (2000).
In this paper we present the other 16 clouds of the sample.
The clouds are distributed along the CMZ, from the Sgr E region
to the vicinity of Sgr D and the ``Clump 2" complex.
Four  clouds are located in the Sgr C complex, three in the vicinity
of Sgr A (two are in  the radio Arc). Two clouds
are situated in the cold dust ridge reported by Lis $\&$ Carlstrom (\cite{LC94})
that seems to connect the radio Arc and Sgr B. Other three clouds belong to
the Sgr B complex.

This  paper is organized as follows. In Sect. 2 we present \CuoO\ and
\utCO\   IRAM-30m observations and        \Hd\ ISO observations.
The analysis of the CO isotopes and \Hd , is presented in
Sect. 3 and 4, respectively.
The results and the possible heating mechanism of the warm gas
are discussed in Sect. 5. 

\section{Observations}
\subsection{IRAM 30-m observations and results}
\begin{figure*}[p]
   \centerline{
       \psfig{figure=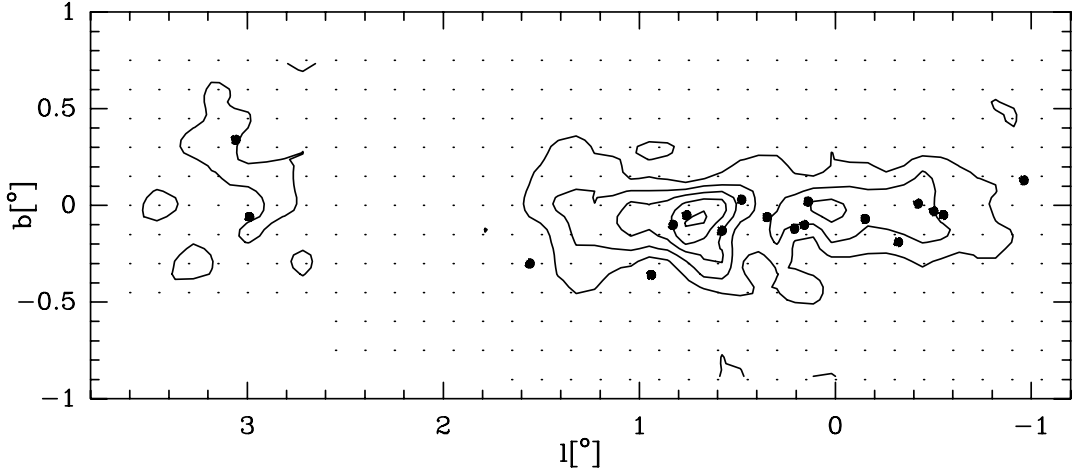,bbllx=27pt,bblly=27pt,bburx=562pt,bbury=283pt,width=15cm}
   }
\caption[]{ The positions of all the sources of our sample
            (including the two clouds presented 
            in Rodr\'{\i}guez-Fern\'andez et~al.~\cite{RF00}) overlayed
            in the \CuoO (1--0) map by Dahmen et al. (1997).} 

  \label{fig_nub}
\end{figure*}

We have observed 
the \juce\  and \jdu\ lines of \utCO\ and \CuoO\ 
with the IRAM 30-m telescope (Pico de Veleta, Spain)
towards the GC molecular clouds given in Table~\ref{tab_pos}.
This table also gives the pointing positions
and the complexes where the clouds belong.
Figure \ref{fig_nub} shows the position of the
sources overlayed on the large scale \CuoO(1$\rightarrow$0)
map of Dahmen et al. (\cite{Dahmen97}).
The observations were carried out in May 1997,  May 1998
and June 2000. 
The \juce\  and \jdu\  lines were observed simultaneously, 
with two 512$\times$ 1 MHz channel filter banks connected to 
two SIS receivers at 3 and 1.3 mm. The receivers were tuned to single side
band mode.
The image rejection, checked against standard calibration sources,
was  always larger than 10 dB.
Typical system temperatures were $\sim 250$~K for the \juce\ lines
and $\sim 400$~K for the \jdu\ lines.
The velocity resolution obtained with  this configuration was
2.7 and 1.4 \kms\ at 3 and 1.3 mm respectively. 
The beam size of the telescope was $22^{''}$ for the \juce\  lines
and $11^{''}$ for the \jdu\   line. 
Pointing and focus were monitored regularly. 
The pointing corrections were never larger than $3^{''}$.
The spectra were taken in position switching with a fixed reference position
at  $(l,b)$=(0\fdg 65, 0\fdg 2), which was selected from the \utCO\ 
map of Bally et al. (\cite{Bally87}).
Calibration of the data was made by observing hot and cold loads
with known temperatures, and the line intensities were converted to
main beam brightness temperatures, $T_\mathrm{MB}$, using main beam efficiencies
of 0.68 and 0.41 for  3 and 1.3 mm respectively. 
The main beam efficiencies for the observations of June 2000
are 0.80 and 0.53 for  3 and 1.3 mm respectively.

\begin{figure*}[p]
  \centerline{
   \psfig{figure=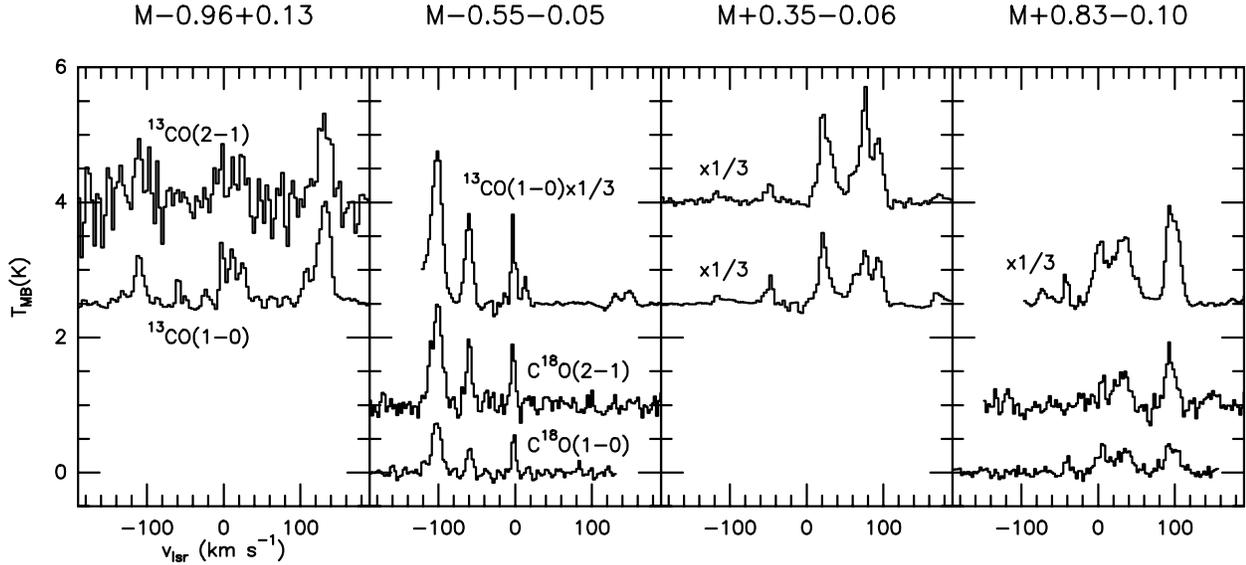,bbllx=44pt,bblly=315pt,bburx=510pt,bbury=548pt,width=17cm}
  }
  \caption[]{\utCO~ and \CuoO~  spectra of four sources.}
  \label{fig_co}
\end{figure*}

A sample of spectra is shown in Fig.~\ref{fig_co}. 
Most of the sources show    CO emission in several velocity
components with Gaussian line  profiles.
However, in some clouds the different components are blended, 
giving rise to more complex profiles.
The observed parameters derived  from Gaussian fits are listed
in Table~\ref{tab_co}.

\subsection{ISO observations and results}
Several \Hd\ pure-rotational lines (from S(0) to S(5))
have also been observed 
towards the  molecular clouds given in Table~\ref{tab_pos}.
The observations were
carried out with the {\em Short Wavelength Spectrometer} (SWS; de Graauw
et~al. \cite{dG96}) on board  ISO. 
The sizes of the SWS apertures at each wavelength are listed in
Table \ref{tab_h2}.
The orientation of the apertures  on the sky varies from source to source,
but it is within position angle 
89.34$^{\circ}$ and 93.58$^{\circ}$ for all the observations
(measuring the angles anti-clockwise between north
and the short sides of the apertures).

The observations presented in this paper are the result of two 
different observing proposals. 
In one of them 
only the S(0), S(1) and S(3) lines were observed,
in the second one all the lines from the S(0)
to the S(5) but the S(2) were observed.
The wavelength bands were scanned in the SWS02 mode  
with a typical on-target time of 100~s.
Three sources were also observed in the SWS01 mode
but the signal-to-noise ratio of these observations is rather poor
and will not be discussed in this paper.
Data were processed interactively at the MPE 
from the Standard Processed Data 
(SPD) to the Auto Analysis Results (AAR) stage
using calibration files of  September 1997
and  were reprocessed automatically through version 7.0 of the
standard Off-Line Processing
(OLP) routines to the AAR  stage.
The two reductions give similar results. 
In this paper we present the results of the reduction with OLP7.0.
The  analysis has been made using
the ISAP2.0\footnote {The ISO Spectral Analysis Package (ISAP) is a joint
                   development by the LWS and SWS Instrument Teams and
                   Data Centers. Contributing institutes are CESR, IAS,
                   IPAC, MPE, RAL and SRON.}
software package. 
With ISAP we have zapped the bad data points and averaged the
two scan directions for each of the 12 detectors. 
Then, we have  shifted (flatfielded) the different detectors to a common
level using the medium value as reference and finally, we have averaged
the 12 detectors and rebinned to one fifth of the instrumental resolution.
No defringing was necessary since the continuum  flux at these 
wavelengths ($\lambda <30 \mu$m) is lower than  30 Jy for all the clouds.

\begin{figure*}[p]
  \centerline{
  \psfig{figure=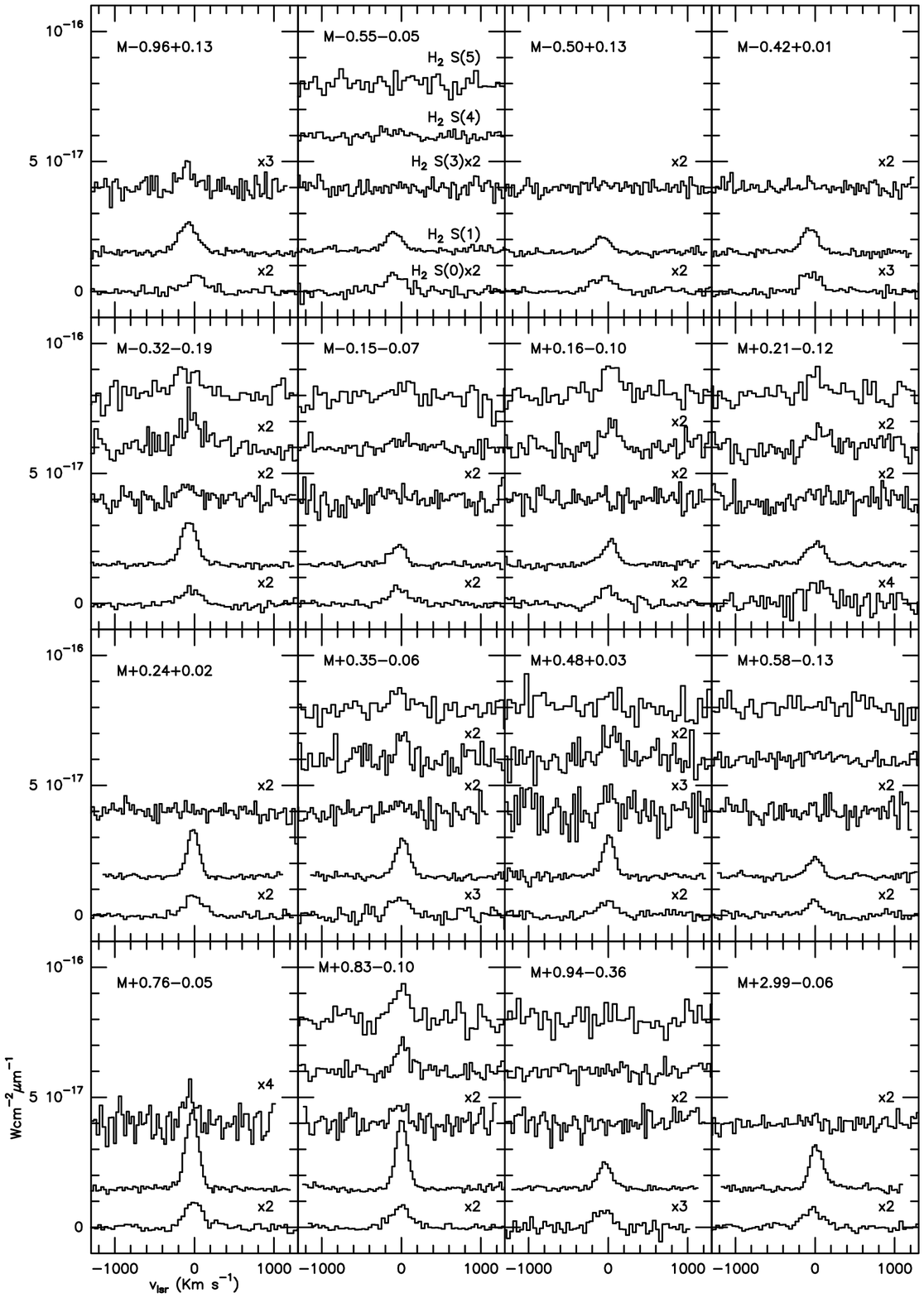,bbllx=30pt,bblly=82pt,bburx=519pt,bbury=788pt,width=17cm}
       }
  \caption[]{\Hd~  spectra.
               They have been rebinned to one fifth of the instrumental
               resolution for point sources.}
  \label{fig_h2}
\end{figure*}

Baseline (order 1) and Gaussian fitting to the lines have also
been carried out with ISAP.
The spectra are shown in Fig. \ref{fig_h2} and the observed fluxes as derived
from  the  fits  are listed in Table \ref{tab_h2}.
The absolute flux calibration errors are less than  
30, 20, 25, 25, and 15$\%$ for the 
S(0), S(1), S(3), S(4), and S(5) lines, respectively (Salama et al. 
\cite{Salama97}).
Because of the medium spectral resolution of the SWS02 mode 
($ \lambda/ \Delta\lambda \sim 1000-2000$) and the 
wavelength calibration uncertainties 
($\sim 15-50$~\kms\ depending on the wavelength,
see  Valentijn et al. \cite{Valentijn96}), it is difficult   
to undertake a detail comparison between  the kinematics of the
\Hd\ lines and those of the \utCO\ and   \CuoO\ lines.
Table \ref{tab_h2} lists the
radial velocities of the S(1) lines, which have the
higher signal-to-noise ratio.
Within the calibration uncertainties,
the radial velocity of the \Hd\ lines agrees  with at least one of 
the  \utCO\ components listed in Table \ref{tab_co}. 

Unfortunately, the lack of resolution does not allow us to
establish if the \Hd\ emission is indeed
arising from  just one or several of the  CO velocity components 
since,  in general, all of them are within 
the velocity range of the unresolved \Hd\ emission.
\ns\ is  the only cloud  for which we can say that the warm
\Hd\ is not likely to arise in all the velocity components seen
in CO. The CO components are centered at -110, 11, and 133~\kms ,
while the \Hd\ S(1)
line is centered at -70~\kms . Even with the spectral
resolution of the SWS02 mode, one can see    that the CO component
with {\it forbidden} velocities (133~\kms ) is not likely to 
contribute   to the \Hd\ emission.

Table \ref{tab_h2} also lists the  widths of   the \Hd~S(1) lines.
The \Hd~ line widths of the GC clouds tend to be larger than the 
instrumental resolution for extended sources ($\sim 170$ \kms~ for the 
S(1) line, see Lutz et al. \cite{Lutz00}).
This is due to the large intrinsic line widths typical of the GC clouds and
mainly,  to the presence of several velocity components along 
the line of sight that  contribute to the \Hd\ emission.
However, not all the sources that show CO emission
in several velocity components have
line widths larger than $\sim 170$ \kms\ (for instance \ot\ or \us).
This implies that not all the CO velocity components detected in these
sources  are contributing to the \Hd\ emission, although it is 
difficult to discriminate which ones are  emitting in \Hd.

   \begin{table}[p]
      \caption[]{J2000 coordinates of the sources}
      \label{tab_pos}    
      \begin{tabular}{llll}
         \hline
         Source&RA&DEC&Complex\\                   
                &h m s & $^\circ$ $^{'}$ $^{''}$ &  \\ 
         \hline
         \ns &17:42:48.3 &-29:41:09.1   &Sgr E \\              
         \cc &17:44:31.3 &-29:25:44.6   &Sgr C \\
         \cce &17:44:32.4&-29:22:41.5   &Sgr C\\       
         \qd &17:44:35.2 &-29:17:05.4   &Sgr C\\       
         \td &17:45:35.8 &-29:18:29.9   &Sgr C\\
         \uc &17:45:32.0 &-29:06:02.2   &Sgr A\\
         \us &17:46:24.9 &-28:51:00.0   &Arc\\
         \du &17:46:34.9 &-28:49:00.0   &Arc \\
         \dq &17:46:07.9 &-28:43:21.5   &Dust Ridge\\
         \tc  &17:46:40.0 &-28:40:00.0  & \\
         \qo  &17:46:39.9 &-28:30:29.2  &Dust Ridge \\
         \co  &17:47:29.9 &-28:30:30.0  &Sgr B \\
         \sis  &17:47:36.8 &-28:18:31.1 &Sgr B\\
         \ot  &17:47:57.9 &-28:16:48.5  &Sgr B \\
         \nq   &17:49:13.2 &-28:19:13.0 &Sgr D \\
         \dnn  &17:52:47.6 &-26:24:25.3 &Clump 2\\
         \hline
      \end{tabular}                      
   \end{table}
\begin{table*}[p]
  \caption[]{Observational parameters and LVG results for  the CO data:
             Integrated intensities of the \juce\  transitions of
             \CuoO\  and \utCO\ and \CuoO\ \jdu\    to  \juce\   line
              intensity ratio. Column densities and \nHd\ derived from the
              LVG calculations.  \NHd\ derived from \NutCO\ assuming a
              \utCO\ abundance relative to \Hd\ of 5 10$^{-6}$.
              Numbers in parentheses are 1$\sigma$ errors
              of the last significant digit.}
  \centerline{
\psfig{figure=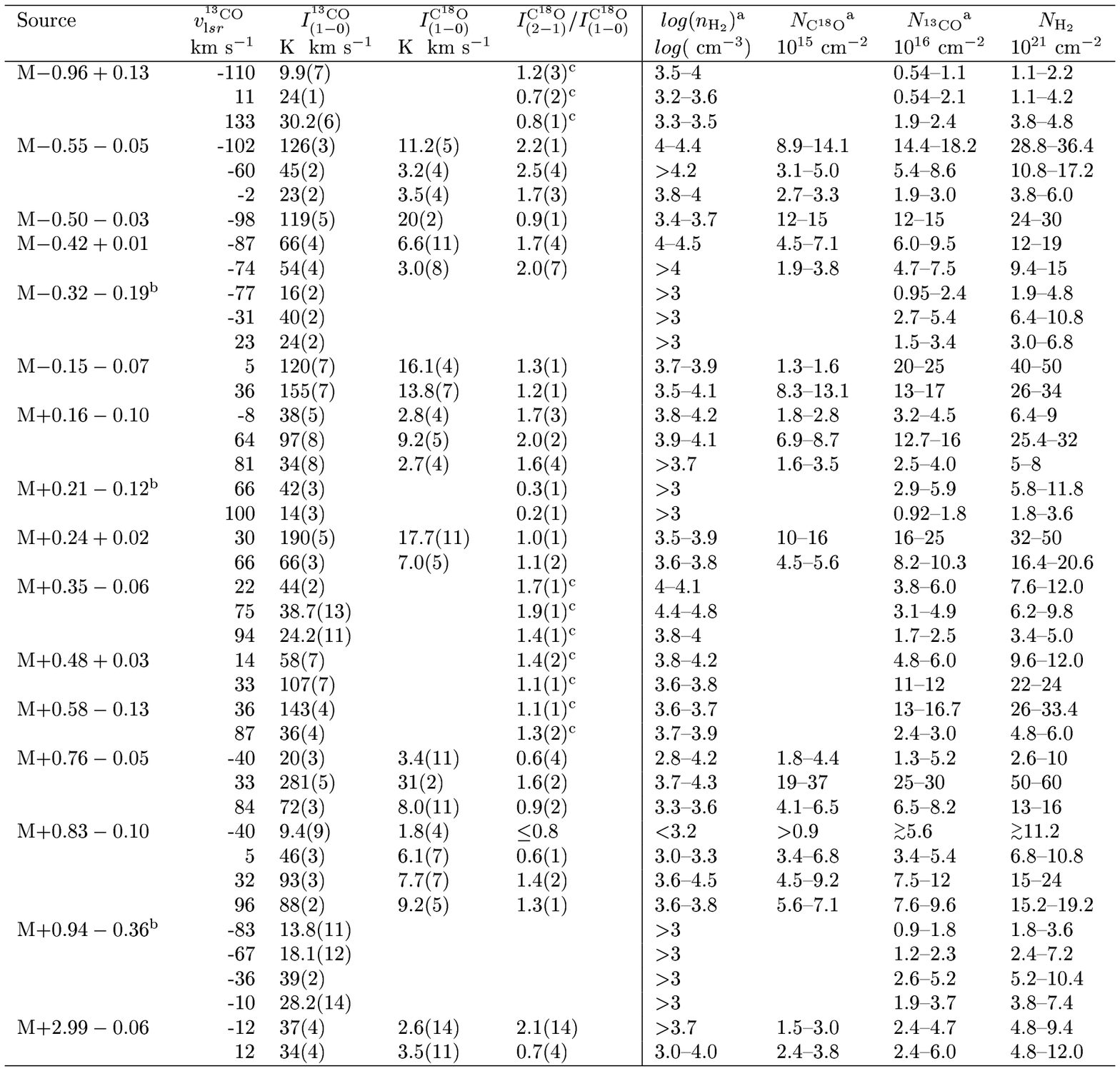,bbllx=5pt,bblly=160pt,bburx=570pt,bbury=730pt,width=17cm}
      }
  \label{tab_co}
  \begin{center}
  \begin{list}{}{}
          \item[$^{\mathrm a}$] Minimum and maximum values derived with
                  the LVG code taking into account the $1\sigma$ errors in
                  the line intensities.
          \item[$^{\mathrm b}$] When the ratio 2-1/1-0 could not be obtained,
               we have assumed \nHd$\geq  10^{3}$ \cmmt\  following
                 H\"uttemeister et al. \cite{Huttemeister98}.
           \item[$^{\mathrm c}$]  $I_{(2-1)}^{\utCO}/I_{(1-0)}^{\utCO}$
                         instead of  $I_{(2-1)}^{\CuoO}/ I_{(1-0)}^{\CuoO}$
  \end{list}
  \end{center}

\end{table*}                                      
   \begin{table*}[p]
       \caption[]{Fluxes of the \Hd\ lines as derived from Gaussian fits
                  in units of 10$^{-20}$ W \cmmd . Upper limits are 3$\sigma$
                  values at the instrumental spectral resolution for point 
                  sources.
                  Numbers in parentheses
                  are 1$\sigma$ errors of the last significant digit 
                  as derived from the Gaussian fits. 
                  The radial velocities and the widths of the lines
                  with better signal-to-noise ratio (the S(1) lines)
                  are also shown. The errors in the radial velocities
                  are dominated by the wavelength calibration uncertainties
                  (15-30 \kms~ for the S(1) line). Typical 1$\sigma$ error
                  of the line widths derived from the Gaussian fits is
                  less than 5 \kms.}

       \label{tab_h2}
       \begin{center}
       \begin{tabular}{llllll|rc}
       \hline             
       \multicolumn{1}{l}{Line} &
       \multicolumn{1}{c}{S(0)   } & \multicolumn{1}{c}{S(1)   } &
       \multicolumn{1}{c}{ S(3)   } &
       \multicolumn{1}{c}{S(4)  } & \multicolumn{1}{c|}{ S(5)  }&
       \multicolumn{1}{c}{$v_\mathrm{S(1)}$} & 
           \multicolumn{1}{c}{$\Delta v_\mathrm{S(1)}$}\\
        Aper. ($~^{''}\times ~^{''}$) & $ 20\times 27$  & $14\times 27$ &
              $14\times 20$ & $14\times 20$ & $14\times 20$& \kms &\kms\\
            $\lambda (\mu$m) &28.2188 & 17.03483  & 9.66491& 
                                             8.02505 & 6.9095& & \\
       \hline
       \ns  & 7.8(9)   & 18.4(8)& 2.2(5)& --& --&-70&270  \\
       \cc &  9.5(14)& 9.7(6)  & \le 0.80 & \le 0.78 & \le 2.0&-80 &230  \\
       \cce &8.2(10)& 8.4(4)  & \le 0.64 & -- & --& -60 & 230 \\
       \qd  & 6.2(6) &13.1(7)  &\le0.70 &-- &--&-57&230 \\
       \td  &7.8(6) & 23.0(6)  & 2.1(2) & 3.5(7) & 5.7(8)&-59 &230 \\
       \uc  &9.4(13) &  9.9(12)  & \le 1.1  & \le 1.4 & \le 2.8 &-35 &220\\
       \us &6.1(9) &10.5(7)  &\le 0.9  & 2.7(6) & 6.5(10)&40 &180\\
       \du &4.7(9) &13.3(8)   & \le 1.2    
           & 2.8(4)$^{\mathrm a}$ & 4.8(11)$^{\mathrm a}$ &16 &260 \\
       \dq & 9.8(5) &18.9(4)  & \le 0.92 & -- &-- & -6 &170\\
       \tc &5.3(8) &17.2(6)  & \le 1.0  & 2.0(7) & 3.5 (8)&27&200 \\
       \qo & 6.7(8)& 15.9(8)  &1.6(3)$^{\mathrm a}$ 
                             &2.4(10)$^{\mathrm a}$ & \le 3.4& 17 &170 \\
       \co & 6.0(6) &8.7(7)  &\le 0.96 & \le 0.97 &\le 2.1& 4 &210 \\
       \sis & 12.4(9) &32.8(8)  & 2.0(5) & -- & -- &-18 &180\\
       \ot & 10.8(9) & 27.1(4)  & 2.2(3) & 5.6(10) & 6.7(8)&16 &170 \\
       \nq & 5.7(9) & 10.6(5)  & \le 1.2  & \le 1.1  & \le 2.7&-30 &190 \\
       \dnn & 9.2(6) &19.4(8) & \le 0.79 & --&-- &28 &190\\
       \hline 
     \end{tabular}  
     \begin{list}{}{}
     \item[$^{\mathrm a}$] Detections with low signal-to-noise ratio 
              ($\sim 2.5$)
     \end{list}
     \end{center}
   \end{table*}      

\section{\CuoO\ and \utCO\ column densities}
The excitation analysis of the three lowest \CuoO\  rotational lines by
H\"uttemeister et al. (1998) shows that  
the CO emission could  arise in cold (20-30 K) and dense ($10^{4}$ \cmmt )
gas or warmer (\gsim 100 K) and less dense gas ($10^{3}$ \cmmt ).
However, the large column densities of cold ($\sim 25$~K) dust
(Mart\'{\i}n-Pintado et al. 1999; Rodr\'{\i}guez-Fern\'andez et al. 2000)
suggest that most of the \juce\ and \jdu\ CO emission should arise from
cold and dense gas coupled to the  dust.

We have derived physical conditions and gas column densities from
the \CuoO\ and \utCO\ data using the Large Velocity Gradient (LVG)
approximation (see e.g. H\"uttemeister et al. 1998). 
Assuming a kinetic temperature (\TK) of 20~K,
we have constrained the \Hd\ densities (\nHd) from the \jdu\
to \juce\   \CuoO\ ratio (or the same  ratio of \utCO\
for a few sources, see Table~\ref{tab_co}).
Then, we have derived the \CuoO\ and \utCO\
column densities (\NCuoO\ and \NutCO ) for corresponding \nHd ,
using the \juce\   lines intensities.
The results of the analysis are listed in Table \ref{tab_co}. 
Typical \CuoO\ \jdu\    to \juce\ line ratios are $\sim$ 1.0--1.5 which give 
\nHd\ of $\sim 10^{3.5-4.0}$ \cmmt\ for \TK=20~K. 
The typical integrated intensities of the \CuoO (\juce )  
lines ($\sim$ 3--9 K \kms) imply  \NCuoO\ of  $\sim 2-8~10^{15}$ \cmmd .
 \NutCO\ is approximately a factor of 10 larger than \NCuoO . 
Since the expected abundance ratio of the
two species in the GC is 12.5 (Wilson \& Matteucci \cite{WM92}), 
the observed ratio indicates that both lines are optically thin.
We can also explain the observed CO lines ratios and intensities
with higher kinetic temperatures (see above). For instance, for
\TK=100 K  one would obtain \Hd\ densities which are 
lower   by a factor of $\sim 2.5$. 
However, even in the unrealistic case that all the \juce\ and \jdu\ CO
arise in warm gas with \TK=100~K, 
the column densities do not change more 
than 10$\%$ with respect to those at low temperature. 
Thus, in general, if one  considers a mixture of warm and cool gas the 
total column densities
traced by CO will be similar to those derived with \TK=20~K.

Table \ref{tab_co} also gives the estimated \Hd\ column densities as
derived from  \NutCO\ assuming that the abundance of \utCO\ relative to
\Hd\ is 5 10$^{-6}$. This ratio is based on the   
$^{13}$C/$^{12}$C   isotopic ratio in the Galactic center of  1/20 
(Wilson \& Matteucci \cite{WM92}) and a CO/\Hd\  ratio of $10^{-4}$
(see e.g. H\"uttemeister et al. \cite{Huttemeister98} and references
therein).     
The typical \Hd\ column densities derived for the main velocity components
in all the sources
are  of a few $10^{22}$  \cmmd. 
\section{Warm \Hd}
Table \ref{tab_h2} lists the observed fluxes of the  \Hd\ lines. 
The most intense
lines are the S(0) (\jdce) and S(1) (\jtu) lines, with typical 
fluxes of 0.5--1 10$^{-19}$ and 1--2 10$^{-19}$  W \cmmd, respectively.
Unfortunately, the S(2) line was only observed in  the two clouds 
already discussed in detail by Rodr\'{\i}guez-Fern\'andez et al. (\cite{RF00}).
The S(3) line is very weak and it has only been detected 
in the   sources with more intense S(1) emission.
Even in some sources which show emission in the S(4) and S(5) lines,
the S(3) line has not been detected. 
This is  due to  strong dust absorption  produced by the solid state
band of the silicates at $9.7\mu$m (Mart\'{\i}n-Pintado et al. \cite{MP99}).

The  pure rotational lines of \Hd~ arise due to electric quadrupole transitions.
The quadrupole transition probabilities are small 
(Turner et al. \cite{Turner77})
and thus the rotational lines remain optically thin. 
In this case,  the column density of the upper level involved
in a transition  from level $i$ to level  $j$  can be  obtained from the
line fluxes $ F_{ij} $  of Table \ref{tab_h2} using the following expression:

\(N_i=\frac{\lambda_{ij}}{hc} 
              \frac{4\pi F_{ij} e^{\tau_{ij}}}{\Omega_{ij}A_{ij}}\)

where  $\lambda_{ij}$ and $ A_{ij }$ are the wavelength 
and the quadrupole  probability of the transition, and   
 $\tau_{ij}$   and $\Omega_{ij}$  are the dust opacity and
the  aperture  at  $\lambda_{ij}$, respectively.
Since the column densities are averaged on the  ISO apertures,
in the case of extended sources (assumed homogeneous),
it is not  necessary  to apply any 
additional correction  to account for the different  ISO 
apertures (see also Rodr\'{\i}guez-Fern\'andez et al. 2000).

\subsection{Extinction and ortho-to-para ratio}
\begin{figure}[p]
  \centerline{
    \psfig{figure=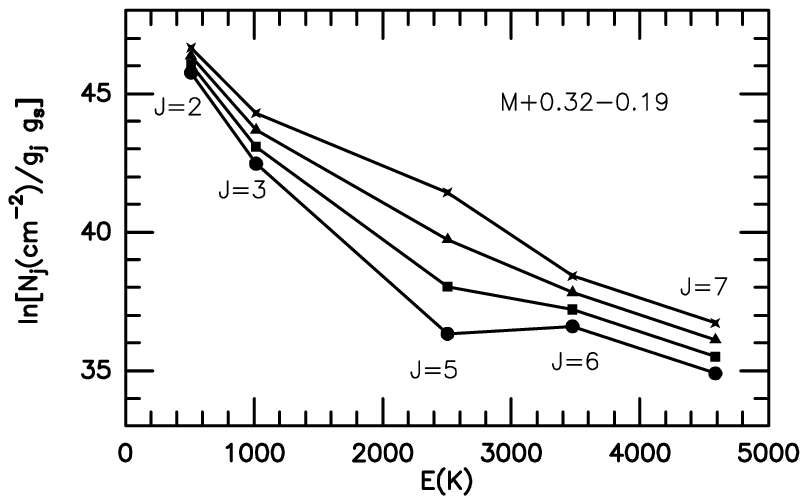,bbllx=35pt,bblly=77pt,bburx=280pt,bbury=246pt,width=8.3cm }
  }
  \caption[]{Population diagrams for \td\ without any extinction
             correction (circles) and corrected for 15 (squares), 
             30 (triangles), and 45 mag. (stars) of visual extinction 
             ($A_{\mathrm V}$). 
             We have assumed the relative extinctions derived toward
             the Galactic center by Lutz (1999). 
             Note that a smooth curve is obtained 
             with $A_{\mathrm V}\sim 30$ mag.}

  \label{fig_bol}
\end{figure}

Figure \ref{fig_bol} shows the population diagrams for one
of the sources for  which more than four lines were detected:
 \td.
It shows, for each observed line, the logarithm of the upper
level population divided by both the rotational  
 and nuclear spin degeneracy,
 i.e.  3($2J+1$) for the ortho levels (odd $J$) and ($2J+1$) for the 
para levels (even  $J$).

The filled circles show the populations 
 without any extinction correction.
One can  see that the population in the $J$=5 level is lower
than expected from the interpolation from the other levels. 
As discussed in Rodr\'{\i}guez-Fern\'andez et al.  (\cite{RF00}), 
this fact can be used to estimate the total extinction 
caused by the dust located between the observer and the \Hd\ 
emitting region.
Once an extinction law is assumed,
we can correct  the \Hd\ line  intensities
by increasing the visual  extinction 
until the column density in any level (in particular that in the $J$=5 level)
is consistent with the column densities derived for  the other 
levels, i.e. until the population diagrams are smooth curves.             

We have used the extinction law derived by Lutz~(\cite{Lutz99}) 
towards the Galactic center using Hydrogen recombination lines.
This   extinction law differs from that of Draine (\cite{Draine89}) 
for silicate-graphite mixtures of grains
in that there is no  deep minimum at $\sim 7 \mu$m and there is 
a slightly higher value for the $A_{9.7 \mu{\mathrm m}}/A_{\mathrm V}$ ratio,
where A$_{\mathrm V}$  the visual  extinction (at 0.55 $\mu{\mathrm m}$)
and $A_{9.7 \mu{\mathrm m}}$ is the extinction at 9.7 $\mu{\mathrm m}$.
For instance, in the case of \td\ one sees  that 15 mag  
of visual extinction (squares in Fig. \ref{fig_bol}) is a lower limit
to the extinction 
while 45 mag (stars in Fig.~\ref{fig_bol}) is an upper limit. 
The best result is obtained for a visual extinction of 
around 30 mag (triangles). 
Using this method  for  the other sources with more than four 
lines detected, we 
also derive  a visual extinction of  $\sim 30$.  
This value should be considered  as a lower limit to the actual extinction
for the sources where the S(3) line was not detected.
  It is not not possible to know how much of this extinction is caused
by material in the line-of-sight towards the GC ({\em foreground extinction})
and how much  is intrinsic to the GC clouds. 
Nevertheless, a visual extinction of $\sim 30$ mag is 
in agreement  with the  average {\em foreground extinction} 
as measured by Catchpole et al. (\cite{Catchpole90}) using  
stars counts and suggests that  the \Hd\ emission can  arise 
from the clouds surfaces (see also Pak, Jaffe \& Keller \cite{Pak96}). 
In the other sources where we cannot estimate the extinction
from our \Hd\ data we have  applied a correction of $A_{\mathrm V}$=30 mag. 
For those clouds   located farther from the center of the Galaxy 
and/or the Galactic plane,  we have corrected the observed fluxes
by 15 mag (see Table \ref{tab_resh2}). This value was  derived 
by Rodr\'{\i}guez-Fern\'andez et al.  (\cite{RF00}) by 
analyzing the far infrared dust emission toward two sources in
the ``Clump 2" and the {\em l=1.5$^{\circ}$} complexes. 
In any case, the extinction correction has a minor
impact in the main results of this paper (see below).
Figure~\ref{fig_ed} shows the extinction corrected population
diagrams for all the sources presented in this paper.

   The values of  extinction required to give   a smooth population diagram
would be somewhat smaller if the \Hd\ ortho-to-para (OTP) ratio
were lower than the local thermodynamic equilibrium (LTE) ratio. 
This is obvious since the 
method to derive the extinction depends mainly 
on the extinction at the wavelength of an ortho level ($J$=5).
   Non-equilibrium OTP ratios measured with the lowest rotational lines
has been found in two clouds of our sample 
(Rod\'{\i}guez-Fern\'andez et al.  2000).
   Unfortunately,  for the clouds presented in this paper,
it is difficult to estimate  the OTP ratio
since the S(2) line has not been observed
and the S(3) line is completely extincted in most of them.   
Current  data do not show any evidence for a non-equilibrium
OTP ratio, but we cannot rule it out {\it a priori}.
For instance, assuming OTP ratios of $\sim 2$ 
we still  can find a smooth 
population diagrams, i.e.  without 
the typical zig-zag shape characteristic   of non-equilibrium OTP ratios
(see. e.g. Fuente et al. 1999).
In this case, the extinction would  be of  $\sim 20-25$~mag instead of 30~mag.
On the contrary, assuming OTP ratios of $\sim 1$
one finds, in general, rather artificial diagrams, which  suggests that 
OTP ratios as low as $\sim 1$ are not compatible with the data.
   Although one must bear in mind these considerations, in the following
we  assume that the OTP ratios are LTE.

\begin{figure*}[p]
  \centerline{
   \psfig{figure=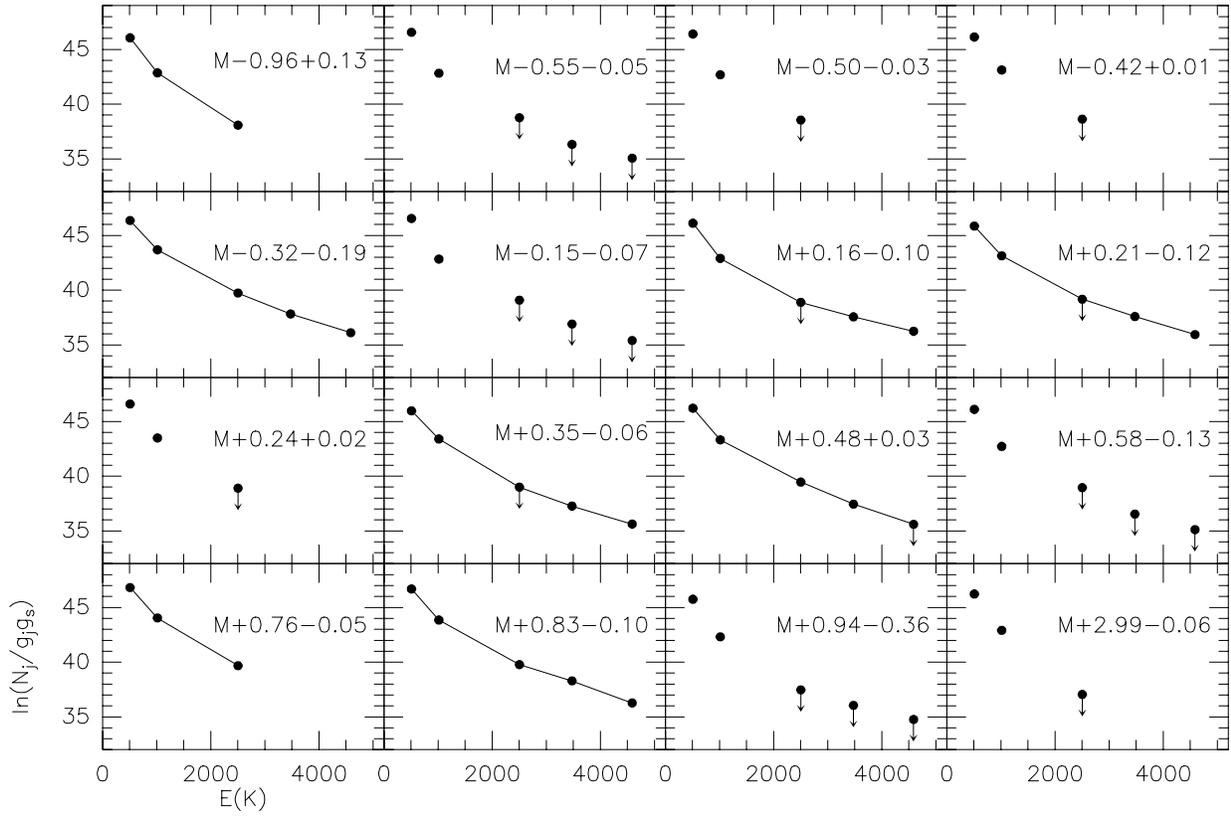,bbllx=15pt,bblly=30pt,bburx=558pt,bbury=393pt,width=17cm}
    }
  \caption[]{Population diagrams for all the sources corrected for
          the extinctions listed in Table \ref{tab_resh2}. 
          The  filled circles
          are connected when more than three lines are detected.
          Arrows indicate upper limits. 
          The error-bars  are smaller than the 
           circles (even taking into account both calibration and
               Gaussian fitting errors).          }
  \label{fig_ed}
\end{figure*}    
   \begin{table}[p]
       \caption[]{Total \Hd\ column densities and rotational temperatures 
                   between the $J$=3 and the $J$=2
                   levels ($T_{32}$) and between the $J$=7 and the $J$=6
                   levels ($T_{76}$) after correcting for extinction.
                   Numbers in parentheses are 1$\sigma$ errors of the
                   last significant digits as derived from the fluxes errors
                   in the Gaussian fits of the lines.}
       \label{tab_resh2}
       \begin{center}
       \begin{tabular}{lllcc}
          \hline             
          \multicolumn{1}{l}{Source} & \multicolumn{1}{c}{$A_{\mathrm V}$}& 
          \multicolumn{1}{c}{$T_{32}$} & \multicolumn{1}{c}{$T_{76}$}& 
          \multicolumn{1}{c}{ \NHd ($T_{32}$)}  \\
          \multicolumn{1}{l}{} & \multicolumn{1}{c}{}& 
          \multicolumn{1}{c}{K} & \multicolumn{1}{c}{K}& 
          \multicolumn{1}{c}{10$^{22}$\cmmd}  \\

          \hline

          \ns   & 15 & 157(6) & -- & 1.10(9) \\
          \cc   & 30 & 135(5) & -- & 2.7(3) \\
          \cce  & 30 & 135(4) & -- & 2.3(2)  \\
          \qd   & 30 & 167(6) & -- & 1.03(8) \\
          \td   & 30 & 188(5) & 650(90) & 1.03(5)\\
          \uc   & 30 & 136(6) & --  & 2.6(4)\\
          \us   & 30 & 157(7) & 900(200) & 1.17(13)\\
          \du   & 30 & 186(13)& 670(110) & 0.64(7)\\
          \dq   & 30 & 163(2) & --       & 1.73(6) \\
          \tc   & 30 & 195(11)& 700(200)  & 0.66(5)\\
          \qo   & 30 & 174(7)& \le 600 & 1.03(9)\\
          \co   & 30 & 149(5) & --       & 1.3(2)\\ 
          \sis  & 30 & 181(4) & --       & 1.77(8) \\
          \ot   & 30 & 178(5) & 550(60)  & 1.59(6) \\
          \nq   & 15 & 146(7) & --     &  0.95(10)\\
          \dnn  & 15 & 152(3) & --       & 1.40(9) \\ 
          \hline
       \end{tabular}
     \end{center}
     \end{table}    

\subsection{\Hd\ column densities and excitation temperatures}

Table \ref{tab_resh2} lists the results derived from the \Hd\ lines
after applying the extinction corrections. 
The excitation temperature derived
from the $J$=3 and $J$=2 levels ($T_{32}$) is between 130 and 
200~K while the excitation temperature derived from the $J$=7
and $J$=6 levels ($T_{76}$) is $\sim 500-700$~K. 
The temperatures are only 15-20 $\%$ larger than those 
one obtains without any extinction correction.
For the four clouds in which the S(4) and S(5)  lines
were undetected, we derive $T_{32}$ of $\sim$ 135--150 K, clearly
lower than the temperature derived for the sources where the
S(4) and S(5)  lines were detected. 
There is no clear dependence of $T_{32}$ on the distance to the
Galactic center. However, it is noteworthy that the two clouds
with lower $T_{32}$ are located in the Sgr C complex, one of them
in the non-thermal filament.

Obviously, $T_{32}$ lacks of physical sense if the ortho-\Hd\
and para-\Hd\ abundances are not in equilibrium.
As mentioned, we can obtain smooth population diagrams
assuming OTP ratios lower than the LTE ratio.
The temperature $T_{32}$ derived in
this case ($T_{32}^\mathrm{corr}$) is higher than the one
derived directly from the observations ($T_{32}$).
For instance, assuming OTP ratios $\sim 2$ one obtains a    
$T_{32}^\mathrm{corr}$ which is $\sim 10\%$ larger than  $T_{32}$.

It is possible to estimate the total warm \Hd\ column densities (\NHd )
by extrapolating the populations in the $J$=2 level to the
$J$=1 and $J$=0 levels at the temperature $T_{32}$. 
The derived warm \NHd\ are listed in Table \ref{tab_resh2} and should
be considered  lower limits to the actual amount of warm molecular gas
since the lowest levels can be populated with colder,
although still warm, gas.
The total column density   of warm \Hd\ varies from source to source
but it is typically of 1--2 10$^{22}$ \cmmd.
These column densities are 
only a factor of 1.2 higher than those one would obtain without any
extinction correction.
Thus, in regard to the derived gas temperatures and 
total column densities, the extinction correction is not critical.
On the other hand, extrapolating the column densities in the
$J$=6 and $J$=7 to lower levels at the temperature $T_{76}$, one finds
that the amount of gas at $\sim 600$~K is  less
than 1$\%$ of the column densities measured at $\sim 150$~K.
The \Hd\ total column densities at temperatures $T_{32}^\mathrm{corr}$
assuming an OTP ratio of $\sim 2$ are lower
than those of Table~\ref{tab_resh2} by a factor of 1.8.
Note, that in  this case the total column density should be 
derived  extrapolating
the {\em observed}  population in the $J$=3 to the $J$=1 level
and the population in the $J$=2 to the  $J$=0 levels,
as two different species at temperature $T_{32}^\mathrm{corr}$.
Of course, these column densities  are still  lower limits 
to the actual warm \Hd\ column densities.

These results are   the first direct estimation of the \Hd\
column densities and the structure of the warm gas in the GC clouds. 
They  show the presence of large column densities
of warm molecular gas with  large temperature gradients (150--700 K),
extending the results derived by H\"uttemeister et~al.~(\cite{Huttemeister93}) 
from their \NHt\ data.

\section{Discussion} 
  
\subsection{Warm \Hd\ to CO and \NHt\ ratios}

As mentioned in Sect. 2.2, we cannot identify which velocity 
components seen in CO are associated to the warm \Hd. 
Furthermore, the bulk of the CO seen in the 
\juce\ and \jdu\ lines do not show the characteristics of 
warm CO associated to the warm \Hd\ (see Sect. 3).
In the following, we will  estimate  the ratio of the 
warm \Hd\ column densities 
observed with ISO to the \Hd\ column densities  derived from the CO using
LVG calculations. We have  added  
the column densities of each velocity component in every source. 
These total \Hd\ column densities are listed in Table~\ref{tab_dis}.
One can compare the
maximum column densities derived from CO to the
\Hd\ column densities listed in Table \ref{tab_resh2} to derive
a lower limit to the fraction of warm molecular  gas  
with respect to the gas emitting in CO.  
These ratios are given in Table~\ref{tab_dis} for all the molecular clouds.
We find that the warm \Hd\ is about $\sim 30 \%$ of the \Hd\ column
densities measured from CO.
For a few  clouds the fraction of warm gas is as high as
77$\%$ (\cce ) or even $\sim 100 \%$ for  \ns .  
This implies that,  for two  clouds all the CO emission
should arise from  warm gas.


Table~\ref{tab_dis} also lists the \NHt\ abundances in the warm
($X$(\NHt )$_\mathrm{warm}$) and cold components 
($X$(\NHt )$_\mathrm{cold}$).
 The  $X$(\NHt )$_\mathrm{warm}$
has been derived from the  column densities of warm ammonia
(H\"uttemeister et al. 1993) and our warm \Hd\  column densities.
We find that, 
$X$(\NHt )$_\mathrm{warm}$ is within a range of 3~10$^{-8}$ to 4~10$^{-7}$.
On the other hand, $X$(\NHt )$_\mathrm{cold}$ has been derived 
from the cold ammonia column densities of H\"uttemeister et al. 
and the  \Hd\ column densities  derived from the \utCO\ data. 
In this case, we have taken into 
account only the \utCO\ velocity components with \NHt\ emission
and we have assumed that, in average, $\sim 70 \%$ of the gas traced
by CO is cold gas.
With these assumptions,
 $X$(\NHt )$_\mathrm{cold}$  varies   between   4~10$^{-8}$ and  6~10$^{-6}$,
being the average value $\sim 5~10^{-7}$.
This is  similar to the abundance in the warm component, and
approximately 10 times higher than the ``typical" interstellar 
ammonia abundance (Irvine et al. \cite{Irvine87}). 
The high \NHt\ abundances in the cold gas point to the existence of
a cold post-shocked gas component as suggested by
H\"uttemeister et al. (1998) to explain the SiO emission in the  GC clouds.

   \begin{table*}[p]
     \caption[]{Total column densities of \Hd\ derived from \utCO .
		Fraction of warm \NHd\ as measured with ISO to the
		total \NHd\ derived from \utCO . Abundances of \NHt\
		in the warm and cold components (see text).}
     \label{tab_dis}
     \begin{center}
      \begin{tabular}{lllll}
      \hline
      \multicolumn{1}{l}{Source} &
      \multicolumn{1}{l}{\NHd$^\mathrm{CO}$ } & 
      \multicolumn{1}{c}{\NHd$^\mathrm{warm}$/\NHd$^\mathrm{CO}$ }&
      \multicolumn{1}{l}{$X$(\NHt )$_\mathrm{warm}$ } &
      \multicolumn{1}{c}{$X$(\NHt )$_\mathrm{cold}$ } \\
      \multicolumn{1}{l}{} &
      \multicolumn{1}{l}{10$^{22}$ \cmmd } & \multicolumn{1}{c}{ }&
      \multicolumn{1}{l}{ } &
      \multicolumn{1}{c}{ } \\
      \hline
      \ns &  0.6-1.1   &  1   & 3.7 10$^{-7}$  & 4.9 10$^{-6}$  \\
      \cc & 4.3-6.0    & 0.45 &      &    \\
      \cce& 2.4-3.0    & 0.77 & 2.6 10$^{-8}$  & 1.6 10$^{-7}$  \\
      \qd & 2.1-3.4    & 0.29 & 8.3 10$^{-8}$ & 2.9 10$^{-8}$  \\
      \td & 1.1-2.2    & 0.45 & 1.8 10$^{-8}$ & 3.1 10$^{-7}$  \\
      \uc & 6.6-8.4    & 0.31 & 2.4 10$^{-7}$ & 2.7 10$^{-7}$  \\
      \us & 3.7-4.9    & 0.24 &      &      \\
      \du & 0.8-1.5    & 0.41 &      &    \\
      \dq & 4.8-7.1    & 0.24 & 1.3 10$^{-7}$ & 8.9 10$^{-7}$ \\
      \tc & 1.7-2.7    & 0.25 &      & \\
      \qo & 3.2-3.6    & 0.28 &      &\\
      \co & 3.1-3.9    & 0.33 &      & \\
      \sis& 6.6-8.6    & 0.21 &      & \\
      \ot & 4.8-6.5    & 0.25 &      & 3.4 10$^{-8}$\\
      \nq & 1.3-2.9    & 0.33 &      & 6.7 10$^{-7}$  \\
      \dnn & 1.0-2.1   & 0.65 &      & 9.0 10$^{-7}$ \\
      \hline
     \end{tabular}
     \end{center}
     \end{table*}    

\subsection{Heating mechanism}             

What is the heating mechanism that
produces such a large amount of warm molecular gas in the GC?
Shocks have been invoked to explain the widespread distribution
and the large abundances
of refractory molecules like SiO (Mart\'{\i}n-Pintado et al. 1997,
H\"uttemeister et al. 1998), the high temperatures observed in  \NHt\
(Wilson et al. 1982, G\"usten et al. 1985) and the non-equilibrium
\Hd\ ortho-to-para ratio of two sources in our sample
(Rodr\'{\i}guez-Fern\'andez et al. 2000).
The high  \NHt\ abundance derived in the previous section 
points to a mechanical heating mechanism since the ammonia molecule
is easily photo-dissociated by  ultraviolet radiation.
The small column densities of warm dust in these clouds also points
to a mechanical heating mechanism (Mart\'{\i}n-Pintado et al. 1999).

On the other hand, in some of the
clouds we have detected  line emission from ionized 
species like \ion{Ne}{ii}, \ion{Ne}{iii} or \ion{O}{iii}, that
should arise in    an  \ion{H}{ii} region ionized by ultraviolet (UV) photons
(Mart\'{\i}n-Pintado et al. 1999, 2000).
This implies that, at least in those clouds, there must be   a
PDR  in the interface between the \ion{H}{ii}
region and the molecular material.
Large scale emission of the \Hd\ $v$=1--0 S(1) line
has also  been interpreted as arising from  PDRs of density
$n \sim 10^4$ \cmmt\ and incident far-UV flux of
$G_0 \sim 10^3$ (in units of 1.6\,10$^{-3}$ ergs \cmmd\ \smu)
in the clouds surfaces (Pak, Jaffe \& Keller 1996).
The total visual extinction of $\sim$ 30 mag derived for
the clouds of our sample matches the expected
{\it foreground} extinction and
suggest that the pure-rotational \Hd\  emission
could also arise in the surfaces of the clouds as
the ro-vibrational lines.

\begin{figure}[p]
  \centerline{
      \psfig{figure=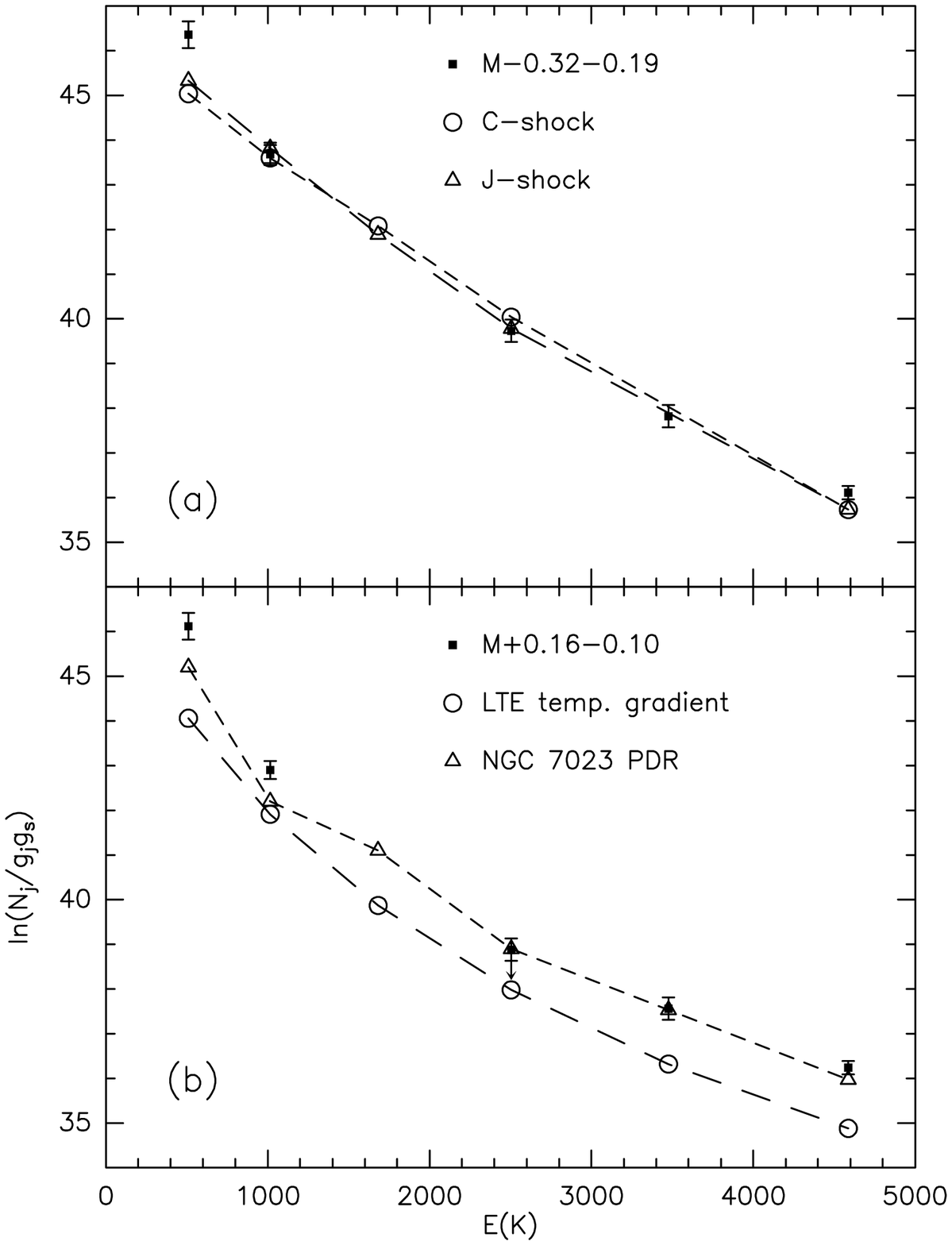,bbllx=35pt,bblly=38pt,bburx=472pt,bbury=620pt,width=8.3cm }
	}
    \caption[]{{\bf a:} Population diagram for \td\ (open squares) 
               corrected for 30 mag. of visual extinction.
	        The errorbars represent upper limits to 
	             the flux calibration uncertainties (see text).
	       For comparison, it also displays the population diagrams
	       derived from the model of  Draine et al. (1983) of a shock
	       with velocity $\sim 12 $\kms and preshock density 10$^6$ \cmmt
	       (circles and dashed lines). Triangles and long-dashed lines
	       are used to plot the population diagram derived from the 
	       J-shock model of Hollenbach \& McKee (1989) for a
	       velocity of 50 \kms and a preshock density of 10$^6$ \cmmt .
	       {\bf b:} Comparison of  the population diagram derived
	       for \us\ (open squares) with the results of 
	       Fuente et al. (1999) for the
	       NGC 7023 PDR (triangles and dashed line) and the population
	       diagram one obtains integrating the H$_2$ emission along
	       the temperature and H$_2$ abundance gradient derived by
	       Burton et al. (1990) for a PDR with density of 10$^6$ \cmmt\
	       and $G_0=10^4$ (open  circles)}
\label{fig_exc}
\end{figure}

We have compared the population diagrams obtained for the GC clouds
with the same type of diagrams predicted by models of C-shocks, J-Shocks 
and PDRs.                 
Figure~\ref{fig_exc}a shows the comparison 
between the predictions of a C-Shock from Draine et al. (\cite{Draine83}),  
a J-Shock from Hollenbach $\&$ McKee (\cite{HM89}), and the data for  \td.
The S(1) to S(5) lines (squares) can be explained with both a C-Shock with
velocity of $\sim 12$ \kms\ acting on gas with preshock density of
10$^6$ \cmmt\ (circles) or a J-Shock of 50 \kms\ and preshock density of
10$^6$ \cmmt\ (triangles).
However, the observed emission in the S(0) line is $\sim 3$ times 
larger than the predicted by both models.

Figure~\ref{fig_exc}b  shows the 
population diagram for \us\ (squares)
versus the prototypical reflection nebula NGC 7023 (triangles).
As discussed by Fuente et al. (1999), the \Hd\ emission from 
this source is well fitted by the PDR model 
of Burton et al. (\cite{Burton90}, \cite{Burton92})
with $G_0=10^4$ and $n$=10$^6$ \cmmt\ although with  an OTP ratio of 1.5-2.
Comparing the NGC 7023 population diagram with \us , 
one finds that the agreement is 
excellent for the S(4) and S(5) lines but it is
not so good for the lowest lines, even taking into account the
non-equilibrium OTP ratio found in NGC 7023. 
In particular, the GC clouds exhibit more emission in the lowest lines
than expected from  the PDR model for $G_0=10^4$ and $n$=10$^6$ \cmmt.
In contrast, the \Hd\ $v$=1--0 S(1) intensity predicted by this
PDR model  is a factor of $\sim 10$ larger than observed by
Pak, Jaffe \& Keller (1996).
This fact would imply that the vibrational  line emission
is more diluted in a 3$^{'}$ 
beam than the pure-rotational lines in the SWS beam or that the PDR
models do not apply.                     

In any case, the observed curvature of the population diagrams
seems to be in good  agreement with the predicted temperature
gradient in a  PDR.  
In Fig.~\ref{fig_exc}b, we also show the population
diagram one obtains integrating the
\Hd\ emission in LTE  
with  the temperature and \Hd\ abundance profiles along
the $G_0$=10$^4$ and $n$=10$^6$ \cmmt\  PDR model of Burton et al. (1990).  
The result differs
from that of Burton et al. in that we do not
take into account any radiative
pumping, which affects mainly to higher levels than those involved in the
S(0) and  S(1) lines. Althoght the GC emission is
$\sim 3$ times larger, it is evident that the shape
of the population diagram is very similar to that observed.

With regard to those sources where the S(4) and S(5) were not detected,
the upper limits imply that if they are PDR-excited
the density must be somewhat lower than $n$=10$^6$ \cmmt , or
if shock-excited, the shock velocity should be slightly lower than
those of the models ploted in Fig.~\ref{fig_exc}.

Both shock and PDR models suggest densities as high as $10^6$ \cmmt\
and fail to explain the observed intensity of the  S(0) emission and
to less extend the S(1) line.
The densities implied by the models seem somewhat large, but
it looks like  the \Hd~ traces two components: a hot ($\sim 500$ K) and
dense (\lsim $10^6$ \cmmt ) component necessary to explain
the observed S(4) and S(5) lines, and a warm component ($\sim 150$ K)
traced by the S(0) and S(1) lines. 
To match the measured  \jdu /\juce\  \utCO\ and \CuoO~  ratios
the warm \Hd~ component should have densities of $\sim 10^3$ \cmmt~
(see Sect. 3).
The hot and dense gas would have \jdu /\juce\ \utCO\ ratios of 
$\sim 4-5$ but it  would emit mainly in the high-$J$ CO lines. 
In any case,  the column density of hot and dense gas  
is very small to make it  detectable in the low-$J$ CO lines
when mixed with the colder and less dense gas that dominates the emission 
of these lines.

To explain the derived $T_{32}\sim 150$~K is necessary to invoke
PDRs with $G_0 \sim 10^3$ and $n \sim 10^3$~\cmmt , but to obtain the
observed intensities $\sim 20$ of such PDRs  are needed. 
J-shock models do not predict temperatures as low as $150$~K.
Moreover, the high velocities required to explain our data are difficult to
reconcile with the observations.
C-shocks could explain the observed S(0) and S(1) emission with,
at least, 10 shocks with velocities as low as $\sim 7$~\kms\ and 
$n$=10$^6$~\cmmt\ (even more shock fronts are needed for lower gas densities).
In addition, dissipation of supersonic turbulence 
could heat the gas to temperatures of $\sim 150$ K 
(Wilson et al. \cite{Wilson82}; G\"usten et al. \cite{Gusten85})
and thus, could contribute to the emission in the two lowest \Hd\ lines.
The origin of  the turbulence would be
the movement of dense clumps in a less dense inter-clump medium
due to the differential Galactic rotation and the tidal 
disruption of the clumps.

The heating rate by dissipation of supersonic turbulence can be estimated
as $\Gamma \sim 3.5\,10^{28} v_{t}^3 n_{\Hd} (1 \mathrm{pc}/l)$ 
erg \smu \cmmt~ (Black \cite{Black87}), 
where $l$ and $v_t$ are  the spatial scale and the velocity of the turbulence,
respectively. 
Taking $v_t\sim 15$ \kms~ (the typical linewidths of GC clouds),  
$l=5$~pc, and $n_{\Hd }=10^3$~\cmmt, one obtains 
$\Gamma \sim 5\,10^{-22}$~erg \smu \cmmt.
For the conditions of the warm gas, $T\sim 150$~K and $n_{\Hd}\sim 10^3$~\cmmt,
the cooling is expected to be dominated by \Hd~ and CO.
Le Bourlot et al. (\cite{LB99}) has recently estimated the cooling rate by \Hd~
($\Lambda_{\Hd}$)
for a wide range of parameters. 
For the warm gas component of the GC clouds we obtain
$\Lambda_{\Hd} \sim 3\,10^{-22}$ erg \smu \cmmt,
which is comparable to the CO cooling rate (see e.g. 
Goldsmith $\&$ Langer \cite{GL78})
Thus, comparing heating and cooling rates, one finds that
the dissipation of supersonic turbulence could account for the 
heating of the warm component.

In summary, several agents could heat the warm component, while
the hot component should trace the densest gas in the GC clouds heated
by a PDR or a  shock.                   
For instance, if the inhomogeneous structure revealed in the 
Sgr B2 envelope by interferometric \NHt\ observations 
(Mart\'{\i}n-Pintado et al. 1999) is common in the GC, and
due to evolved massive stars as they propose, 
both C-shocks of $\sim 10$ \kms\ (shell expansion) and PDRs
(stellar radiation) would be present.
However, it is not possible to rule out mechanical heating 
by large scale shocks.
In fact,      the high fraction of warm \Hd\ derived for \ns\ and the fact
that the CO component with positive velocities apparently does not 
contribute   to the \Hd\ emission suggests this kind of heating since,
at this galactic longitude, shocks are expected at negative velocities 
due to the intersection of $x_1$ and $x_2$ orbits in the 
context of a barred potential (Binney et al. \cite{Binney91}).

%

\section{Summary and conclusions}
We have observed the S(0) to S(5) \Hd\ pure-rotational lines  with the SWS
spectrometer on-board ISO toward a sample of 18 molecular clouds of the
Galactic center region. 
The S(3) line is strongly affected by dust extinction due to the 9.7 $\mu$m
band of the silicates.
After correcting the \Hd\ data for extinction using a self-consistent method,
and assuming that the ortho- and para-\Hd\ populations are in equilibrium
one finds that the S(0) and S(1) lines indicate temperatures of $\sim 150$ K.
Extrapolating to the lowest levels at that temperature, a total \Hd\ column
density of $\sim 1-2\,10^{22}$~\cmmd\ is derived.
This is the first direct estimate of the column density of warm gas in the 
GC clouds. In addition, it shows a complex temperature structure of
the warm gas.

The temperature derived from the S(5) and S(4) levels is $\sim 600$ K in
the sources in which it can be derived,  however the column density
of gas at this temperature is
less than $1\%$ of the column density at $T$=150~K.
Assuming an OTP ratio of  $\sim 2$ the temperatures would be $10\%$ larger
than those derived assuming a LTE OTP ratio, while 
the total \Hd\ column densities at those temperatures  would be 
a factor of  $\sim 1.8$ lower than the column densities derived assuming the 
ortho- and para-\Hd\ populations in equilibrium.
Comparing the \Hd\ warm column densities with the column densities derived
from our CO data by LVG calculations one finds that the 
average fraction of warm  \Hd\
to the gas observed in CO is $\sim 30 \% $.
With our data and the \NHt\ observations of H\"uttemeister et al. (1993) we 
derive relatively high \NHt\ abundances of a few 10$^{-7}$ in both 
the warm and the cold components.         

Several indirect arguments point to shocks as the heating 
mechanism of the warm gas but PDRs may also play a role.
Direct comparison of the \Hd\ data with  PDRs 
and shocks models 
indicate  that the S(4) and S(5) trace the densest gas in the GC clouds
(\lsim 10$^6$ \cmmt) heated in  PDRs or shocks.
Nevertheless, such dense PDRs or shocks  fail
to explain the S(0) and  S(1) lines: several less dense PDRs, low velocity
shocks ($< 10$~\kms) or both
along the line of sight would be needed to explain the observed emission.

The cooling by \Hd~ in the warm component of GC clouds
is comparable to the cooling by CO.
Equating the \Hd~ cooling rate with the heating rate by
dissipation of supersonic turbulence, one finds that this mechanism
could also  contribute to the emission in the two lowest \Hd~ lines.
In one source (\ns ), we have also found some  evidence of large 
scale shocks that should be checked with higher spectral resolution
\Hd\ observations.

\begin{acknowledgements}
We  thank the referee, Rolf  G\"usten, for his useful comments. 
We acknowledge support from the ISO Spectrometer Data Center at MPE,
funded by DARA under grant 50 QI 9402 3. NJR-F, JM-P, PdV, and AF have
been partially supported by the CYCIT and the PNIE under grants
PB96-104, 1FD97-1442 and ESP99-1291-E.
NJR-F acknowledges {\it Consejer\'{\i}a de Educaci\'on y Cultura de la
Comunidad de Madrid} for a pre-doctoral fellowship.
\end{acknowledgements}

\end{document}